\documentclass[12pt]{article}
\usepackage{amsmath}
\def\a{\alpha}
\def\b{\beta}

\def\d{\delta}
\def\e{\epsilon}

\def\g{\gamma}

\def\r{\rho}
\def\s{\sigma}

\def\be{\begin{equation}}
\def\ee{\end{equation}}
\def\arr{\begin{array}{rll}}
\def\ea{\end{array}}
\def\bea{\begin{eqnarray}}
\def\eea{\end{eqnarray}}

\def\N2{$N{=}2$}

\def\>{\rangle}
\def\<{\langle}
\def\+{\dagger}
\def\={\ =\ }

\begin{document}
\renewcommand{\thefootnote}{\fnsymbol{footnote}}
\begin{titlepage}
\setcounter{page}{0}
\vskip 2cm
\begin{center}
{\LARGE\bf $N=4$ $l$--conformal Galilei superalgebra}\\
\vskip 1cm

$
\textrm{\Large Anton Galajinsky and Ivan Masterov\ }
$
\vskip 0.7cm
{\it
Laboratory of Mathematical Physics, Tomsk Polytechnic University, \\
634050 Tomsk, Lenin Ave. 30, Russian Federation} \\
{Emails: galajin@tpu.ru, masterov@tpu.ru}

\end{center}
\vskip 1cm
\begin{abstract} \noindent
An $N=4$ supersymmetric extension of the $l$--conformal Galilei algebra is constructed.
This is achieved by combining generators of spatial symmetries from the $l$--conformal Galilei algebra and those underlying the most general superconformal group in one dimension $D(2,1;\alpha)$. The value of the group parameter $\alpha$ is fixed from the requirement that the resulting superalgebra is finite--dimensional. The analysis reveals $\alpha=-\frac 12$ thus reducing $D(2,1;\alpha)$ to $OSp(4|2)$.
\end{abstract}


\vskip 0.5cm


\end{titlepage}

\renewcommand{\thefootnote}{\arabic{footnote}}
\setcounter{footnote}0

\noindent
{\bf 1. Introduction}\\
\noindent

There are at least two reasons to be concerned about $N=4$ supersymmetric extensions of the $l$--conformal Galilei algebra \cite{NOR,H}. On the one hand, recently
there has been extensive investigation of $N=4$ superconformal many--body mechanics in one dimension aimed at
a microscopic description of the near horizon extreme Reissner--Nordstr\"om black hole (for a review and further references see \cite{FIL}). In that context, nonrelativistic superconformal algebras provide a natural framework for
higher--dimensional generalizations \cite{G2}. At present, $N=4$ is regarded to be the maximum value for which the construction of interacting many--body models in $d>1$ is feasible.
On the other hand, the study of the nonrelativistic version of the AdS/CFT--correspondence has sparked substantial interest
in nonrelativistic superconformal symmetries and their realizations in field theory and mechanics.\footnote{Literature on the subject is rather extensive. For a discussion relevant for this work see \cite{FL}. Some important earlier developments include \cite{PH}--\cite{MS}.}

Focusing on $d=1$, $N=4$ superconformal many--body mechanics based on the supergroup $SU(1,1|2)$, which is the instance relevant for
a microscopic description of the near horizon extreme Reissner--Nordstr\"om black hole \cite{FIL}, one reveals two prepotentials which govern its dynamics \cite{GLP}. They
obey a coupled set of partial differential equations which are incompatible with translation invariance. Because going beyond one dimension implies enforcing
spatial translation symmetry, the construction of interacting $d>1$, $N=4$ superconformal many--body mechanics based upon $SU(1,1|2)$ seems unfeasible.

The most general $N=4$ supersymmetric extension of the conformal group in one dimension is given by the exceptional supergroup $D(2,1;\alpha)$ which is parametrized by a real number $\alpha$.
Its generators are associated with time translations, dilatations, special conformal transformations, supersymmetry transformations and their superconformal partners, as well as with two variants of $su(2)$--transformations. For $\alpha$, $\frac{1}{\alpha}$, $-1-\alpha$, and $-\frac{\alpha}{1+\alpha}$ the associated Lie superalgebras are isomorphic \cite{FSS}.
As was demonstrated in \cite{KL}, the master equations which underlie $D(2,1;\alpha)$ superconformal many--body mechanics admit translation invariant solutions provided $\alpha=-\frac 12$. This hints at the possibility to built an $N=4$ supersymmetric extension of the $l$--conformal Galilei algebra based upon $D(2,1;-\frac 12) \simeq OSp(4|2)$.

The goal of this work is to formulate the structure relations of an $N=4$ $l$--conformal Galilei superalgebra based upon $osp(4|2)$ superalgebra. This is achieved by adding operators which generate spatial symmetries, including accelerations, to the superconformal algebra $osp(4|2)$ and finding a chain of extra bosonic and fermionic generators which are needed in order to close the full superalgebra.

The work is organized as follows. In the next section we construct a representation of the Lie superalgebra associated with the superconformal group $D(2,1;\alpha)$ in terms of differential operators in a superspace parametrized by one temporal and $d$ spatial coordinates along with four real fermions. In Sect. 3 we extend this superalgebra by
spatial symmetry transformations which underlie the $l$--conformal Galilei algebra. An extra set of bosonic and fermionic operators, which are needed in order to close the full superalgebra, is found. Requiring the superalgebra to be finite--dimensional, one gets the restriction on the value of the group parameter $\alpha=-\frac 12$ which reduces $D(2,1;\alpha)$ to $OSp(4|2)$. In Sect. 4 we obtain the structure relations of
an $N=4$ supersymmetric extension of the $l$--conformal Galilei algebra based upon $osp(4|2)$ which constitute the main result of this work. Our spinor conventions are given in Appendix A. The Lie superalgebra associated with the superconformal group $D(2,1;\alpha)$ is exposed in Appendix B. Throughout the paper summation over repeated indices is understood.

\vspace{0.5cm}

\noindent
{\bf 2. A realization of $D(2,1;\alpha)$ in superspace}\\

\noindent

Consider a superspace parametrized by the temporal variable $t$, spatial coordinates $x_i$, $i=1,\dots,d$, and the
fermionic $SU(2)$--doublets $\theta_\alpha$, $\bar\theta^\a$, $\alpha=1,2$, which are complex conjugates of each other ${(\theta_\alpha)}^{*}=\bar\theta^\a$.
On such a superspace one can realize the Lie superalgebra associated with the exceptional supergroup $D(2,1;\alpha)$ by means of the differential operators
\bea\label{d21}
\begin{aligned}
&
H=\frac{\partial}{\partial t}, \qquad K=-t^2\frac{\partial}{\partial t}+2tD+
(1+2\alpha)\theta_{\gamma}\bar{\theta}^{\gamma} \left(\theta_{\beta}\bar{\theta}^{\beta}\frac{\partial}{\partial t}-i\theta_{\beta}\frac{\partial}{\partial\theta_{\beta}}+i\bar{\theta}^{\beta}\frac{\partial}{\partial\bar{\theta}^{\beta}}\right),
\\[6pt]
&
D=t \frac{\partial}{\partial t} +l x_i \frac{\partial}{\partial x_i} +\frac 12 \left(\theta_\alpha \frac{\partial}{\partial \theta_\alpha}+\bar\theta^\alpha \frac{\partial}{\partial \bar\theta^\alpha}\right),\qquad \mathcal{J}_a=\frac{i}{2} {{(\sigma_a)}_\alpha}^\beta \left(\theta_\beta \frac{\partial}{\partial \theta_\alpha} -\bar\theta^\alpha \frac{\partial}{\partial \bar\theta^\beta}  \right),
\\[6pt]
&
Q_\alpha=\epsilon_{\alpha \beta} \left(i\frac{\partial}{\partial \theta_\beta}+\bar\theta^\beta \frac{\partial}{\partial t} \right),\qquad  {\bar Q}^\alpha=\epsilon^{\alpha \beta} \left(i\frac{\partial}{\partial \bar\theta^\beta}+\theta_\beta \frac{\partial}{\partial t} \right),
\\[6pt]
&
I_{-}=-\epsilon_{\alpha \beta} \bar\theta^\alpha \frac{\partial}{\partial \theta_\beta},\quad I_{+}=\epsilon^{\alpha \beta} \theta_\alpha \frac{\partial}{\partial \bar\theta^\beta},
\qquad  I_3=-\frac{i}{2} \left(\theta_\alpha \frac{\partial}{\partial \theta_\alpha}-\bar\theta^\alpha \frac{\partial}{\partial \bar\theta^\alpha} \right),
\\[6pt]
&
S_\alpha=2 \epsilon_{\alpha \beta} \bar\theta^\beta \left(t \frac{\partial}{\partial t}-D\right)-t Q_\alpha+(\alpha+1/2)\left(i\theta_{\alpha}\bar{\theta}^2\frac{\partial}{\partial t}+2\theta_\alpha I_{-}+\bar{\theta}^2\frac{\partial}{\partial\bar{\theta}^{\alpha}}\right),
\\[6pt]
&
\bar S^\alpha=2 \epsilon^{\alpha \beta} \theta_\beta \left(t \frac{\partial}{\partial t}-D\right)-t \bar Q^\alpha+(\alpha+1/2)\left(i\bar{\theta}^{\alpha}\theta^2\frac{\partial}{\partial t}-2\bar{\theta}^{\alpha}I_{+}+\theta^2 \frac{\partial}{\partial\theta_{\alpha}}\right),
\end{aligned}
\eea
which involve two arbitrary parameters $\alpha$, and $l$. Left derivatives are chosen for the odd variables and the Pauli matrices are designated by $\s_a$ (for our conventions see Appendix A).  Note that at this stage $l$ is a real number. Later on, when adding spatial symmetries, it will be constrained to take (half)integer values only.

The operators above generate time translations ($H$), dilatations ($D$), special conformal transformations ($K$), supersymmetry transformations ($Q_\alpha,{\bar Q}^\alpha$), two variants of $su(2)$--transformations ($\mathcal{J}_a$ and $I_{-},I_{+},I_3$), and superconformal transformations ($S_\alpha,{\bar S}^\alpha$). In order to verify that they obey the structure relations of the Lie superalgebra associated with $D(2,1;\alpha)$ (see Appendix B), one has to extensively use the spinor algebra and properties of the Pauli matrices exposed in Appendix A.
As $(H,D,K)$ form the conformal algebra in one dimension $so(2,1)$, which is also a subalgebra in the $l$--conformal Galilei algebra \cite{NOR,H}, its seems natural to try to extend (\ref{d21}) by the operators which generate spatial symmetries including accelerations.

\vspace{0.5cm}

\noindent
{\bf 3. Extending $D(2,1;\alpha)$ by spatial symmetries}\\

\noindent

From the previous work on the $l$--conformal Galilei algebra \cite{NOR,H} it is known how to realize spatial symmetry transformations in a nonrelativistic spacetime parametrized by $t$ and $x_i$
\be\label{extra}
C^{(n)}_i=t^n \frac{\partial}{\partial x_i}, \qquad M_{ij}=x_i \frac{\partial}{\partial x_j}-x_j \frac{\partial}{\partial x_i}.
\ee
In this condensed notation the upper index in braces labels various spatial symmetry generators, $n=0,\dots,2l$, and $l$ is identified with the parameter which enters the generator of dilatations $D$ above. In particular, $n=0$ and $n=1$ correspond to spatial translations and Galilei boosts, while higher values of $n$ describe accelerations. $M_{ij}$ stand for spatial rotations. At this stage, in order to deal with a finite--dimensional algebra, one has to require $l$ to take (half)integer values only \cite{NOR,H}.

It is straightforward to compute (anti)commutators among (\ref{d21}) and (\ref{extra}). The superalgebra does not close unless one introduces the chain of extra bosonic and fermionic operators
\begin{align}\label{extra1}
&
L^{(n)}_{i \alpha}=\theta_\alpha t^n \frac{\partial}{\partial x_i}, && n=0,\dots,2l-1
\nonumber\\[2pt]
&
{\bar L}^{(n) \alpha}_i=\bar\theta^\alpha t^n \frac{\partial}{\partial x_i}, && n=0,\dots,2l-1
\nonumber\\[2pt]
&
P^{(n) \beta}_{i \alpha}=\theta_\alpha \bar\theta^\beta t^n \frac{\partial}{\partial x_i}, && n=0,\dots,2l-2
\nonumber\\[2pt]
&
R^{(n)}_i=\frac 12 \theta^2 t^n \frac{\partial}{\partial x_i}, && n=0,\dots,2l-2
\nonumber\\[2pt]
&
{\bar R}^{(n)}_i=\frac 12 \bar\theta^2 t^n \frac{\partial}{\partial x_i}, && n=0,\dots,2l-2
\nonumber\\[2pt]
&
Z^{(n)}_{i \alpha}=\frac 12 \theta_\alpha \bar\theta^2 t^n \frac{\partial}{\partial x_i}, && n=0,\dots,2l-3
\nonumber\\[2pt]
&
\bar Z^{(n)\alpha}_{i}=\frac 12 \bar\theta^\alpha \theta^2 t^n \frac{\partial}{\partial x_i}, && n=0,\dots,2l-3
\nonumber\\[2pt]
&
W^{(n)}_{i}=\frac 14 \theta^2 \bar\theta^2 t^n \frac{\partial}{\partial x_i}, && n=0,\dots,2l-4.
\end{align}
These operators can be viewed as fermionic partners of the acceleration generators $C_i^{(n)}$. What is more significant,
in order to get a finite--dimensional superalgebra, one has to fix the value of the group parameter
\be\label{a}
\alpha=-\frac 12,
\ee
which reduces $D(2,1;\alpha)$ to $OSp(4|2)$.

Let us explain how the latter restriction and the range of values of index $n$ exposed in the right column in Eq. (\ref{extra1}) come about. It proves sufficient to analyse commutators involving the generator of special conformal transformations $K$. As a pattern, consider the commutator of $K$ and $\bar{Z}_i^{(n)\alpha}$
\be
[K,\bar{Z}_i^{(n)\alpha}]=(n-2l+3)\bar{Z}_i^{(n+1)\alpha}.
\ee
In order to prevent the unbounded growth of the number of $\bar{Z}_i^{(n)\alpha}$--type generators, one has to require the index $n$, which labels various members of the set, to take on the following values $n=0,\dots,2l-3$. This terminates an unlimited proliferation of such generators. Note that a similar situation holds for the original
$l$--conformal Galilei algebra which involves
\be
[K,C_i^{(n)}]=(n-2l)C_i^{(n+1)}.
\ee

As the next step, let us evaluate the commutator of $K$ and $L_{i\alpha}^{(n)}$
\be
[K,L_{i\alpha}^{(n)}]=(n-2l+1)L_{i\alpha}^{(n+1)}-i(1+2\alpha)\epsilon_{\alpha\beta}\bar{Z}_i^{(n)\beta}.
\ee
Focusing on the first term on the right hand side, one concludes that the range of values of the index $n$, which designates different members of the set $L_{i\alpha}^{(n)}$, should be $n=0,\dots,2l-1$.
This prevents the unbounded growth of the number of $L_{i\alpha}^{(n)}$--type generators. Focusing on the second term, consistency requires the restriction (\ref{a}) to be imposed, otherwise two extra $\bar{Z}_i^{(n)\alpha}$--type generators would be proliferated above the upper bound $(2l-3)$ revealed earlier. Further analysis of the superalgebra shows that this pattern is ubiquitous.
For $l=1$ a similar analysis has been presented in \cite{FL}.

\vspace{0.5cm}

\noindent
{\bf 4. $N=4$ $l$--conformal Galilei superalgebra}\\

\noindent

Having fixed the parameters $l$ and $\alpha$ in (\ref{d21}) and the form of the extra generators (\ref{extra}), (\ref{extra1}), we are now in a position to obtain the structure relations of an $N=4$ $l$--conformal Galilei superalgebra based upon $osp(4|2)$. A straightforward although a bit tedious calculation yields
\begin{align}\label{algebra}
&
[H,D]=H, && [H,K]=2D,
\nonumber\\[2pt]
&
[D,K]=K, && [\mathcal{J}_a,\mathcal{J}_b ]=\epsilon_{abc} \mathcal{J}_c,
\nonumber\\[2pt]
&
\{ Q_\a, \bar Q^\b \}=-2 i H {\d_\a}^\b, &&
\{ Q_\a, \bar S^\b \}={{(\s_a)}_\a}^\b \mathcal{J}_a+2iD {\d_\a}^\b+I_3 {\d_\a}^\b,
\nonumber\\[2pt]
&
\{ S_\a, \bar S^\b \}=-2i K {\d_\a}^\b, &&
\{ \bar Q^\a, S_\b \}=-{{(\s_a)}_\b}^\a \mathcal{J}_a+2iD {\d_\b}^\a-I_3 {\d_\b}^\a,
\nonumber\\[2pt]
&
\{ Q_\a, S_\b \}=i\epsilon_{\alpha \beta} I_{-}, &&
\{ {\bar Q}^\a, {\bar S}^\b \}=-i  \epsilon^{\alpha \beta} I_{+},
\nonumber\\[2pt]
& [D,Q_\a] = -\frac{1}{2} Q_\a, && [D,S_\a] =\frac{1}{2} S_\a,
\nonumber\\[2pt]
&
[K,Q_\a ] =S_\a, && [H,S_\a ]=-Q_\a,
\nonumber\\[2pt]
&
[\mathcal{J}_a,Q_\a] =\frac{i}{2} {{(\s_a)}_\a}^\b Q_\b, && [ \mathcal{J}_a,S_\a] =\frac{i}{2} {{(\s_a)}_\a}^\b S_\b,
\nonumber
\end{align}
\begin{align}
& [D,\bar Q^\a ] =-\frac{1}{2} \bar Q^\a, && [D,\bar S^\a]=\frac{1}{2} \bar S^\a,
\nonumber\\[2pt]
& [K,\bar Q^\a] =\bar S^\a, && [H,\bar S^\a] =-\bar Q^\a,
\nonumber\\[2pt]
&
[\mathcal{J}_a,\bar Q^\a] =-\frac{i}{2} \bar Q^\b {{(\s_a)}_\b}^\a, && [\mathcal{J}_a,\bar S^\a]=-\frac{i}{2}
\bar S^\b {{(\s_a)}_\b}^\a,
\nonumber\\[2pt]
&
[I_{-},\bar Q^\a ]=\epsilon^{\alpha \beta} Q_\beta, && [I_{-},\bar S^\a ] =\epsilon^{\alpha \beta} S_\beta,
\nonumber\\[2pt]
&
[ I_{+},Q_\a]=-\epsilon_{\alpha\beta} \bar Q^\b, && [I_{+},S_\a]=-\epsilon_{\alpha\beta} \bar S^\b,
\nonumber\\[2pt]
&
[I_3,Q_\a ] =\frac{i}{2} Q_\a, &&  [I_3,S_\a ]=\frac{i}{2} S_\a,
\nonumber\\[2pt]
&
[I_3,\bar Q^\a] =-\frac{i}{2} \bar Q^\a, &&  [ I_3,\bar S^\a ]=-\frac{i}{2} \bar S^\a,
\nonumber\\[2pt]
&
[I_{-},I_3] =-i I_{-}, &&  [ I_{+},I_3 ] =i I_{+},
\nonumber\\[2pt]
&
[I_{-},I_{+}] =2 i I_{3}, && [H,C_i^{(n)}]=nC_i^{(n-1)},
\nonumber\\[2pt]
&
[D,C_i^{(n)}]=(n-l)C_i^{(n)}, && [K,C_i^{(n)}]=(n-2l)C_i^{(n+1)},
\nonumber\\[2pt]
&
[Q_{\alpha},C_i^{(n)}]=n\epsilon_{\alpha\beta}\bar{L}_i^{(n-1)\beta}, && [\bar{Q}^{\alpha},C_i^{(n)}]=n\epsilon^{\alpha\beta}L_{i\beta}^{(n-1)},
\nonumber\\[2pt]
&
[S_{\alpha},C_i^{(n)}]=-(n-2l)\epsilon_{\alpha\beta}\bar{L}_i^{(n)\beta}, && [\bar{S}^{\alpha},C_i^{(n)}]=-(n-2l)\epsilon^{\alpha\beta}L^{(n)}_{i\beta},
\nonumber\\[2pt]
&
[H,L_{i\alpha}^{(n)}]=nL_{i\alpha}^{(n-1)}, && [D,L_{i\alpha}^{(n)}]=\left(n-l+\frac 12 \right)L_{i\alpha}^{(n)},
\nonumber\\[2pt]
&
[K,L_{i\alpha}^{(n)}]=(n-2l+1)L_{i\alpha}^{(n+1)}, && [\mathcal{J}_a,L_{i\alpha}^{(n)}]=\frac{i}{2}{(\sigma_a)_{\alpha}}^{\beta}L_{i\beta}^{(n)},
\nonumber\\[2pt]
&
\{Q_{\alpha},L_{i\beta}^{(n)}\}=i\epsilon_{\alpha\beta}C_i^{(n)}-n\epsilon_{\alpha\gamma}P_{i\beta}^{(n-1)\gamma}, && \{\bar{Q}^{\alpha},L_{i\beta}^{(n)}\}=n{\delta_{\beta}}^{\alpha}R_i^{(n-1)},
\nonumber\\[2pt]
&
\{S_{\alpha},L_{i\beta}^{(n)}\}=-i\epsilon_{\alpha\beta}C_i^{(n+1)}+(n-2l+1)\epsilon_{\alpha\gamma}P_{i\beta}^{(n)\gamma}, && [I_{-},L_{i\alpha}^{(n)}]=\epsilon_{\alpha\beta}\bar{L}_i^{(n)\beta},
\nonumber\\[2pt]
&
\{\bar{S}^{\alpha},L_{i\beta}^{(n)}\}=-{\delta_{\beta}}^{\alpha}(n-2l+1)R_i^{(n)}, && [I_3,L_{i\alpha}^{(n)}]=-\frac{i}{2}L_{i\alpha}^{(n)},
\nonumber\\[2pt]
&
[H,\bar{L}_i^{(n)\alpha}]=n\bar{L}_i^{(n-1)\alpha}, && [D,\bar{L}_i^{(n)\alpha}]=\left(n-l+\frac 12\right)\bar{L}_i^{(n)\alpha},
\nonumber\\[2pt]
&
[K,\bar{L}_i^{(n)\alpha}]=(n-2l+1)\bar{L}_i^{(n+1)\alpha}, && [\mathcal{J}_a,\bar{L}_i^{(n)\alpha}]=-\frac{i}{2}\bar{L}_i^{(n)\beta}{(\sigma_a)_{\beta}}^{\alpha},
\nonumber\\[2pt]
&
\{Q_{\alpha},\bar{L}_i^{(n)\beta}\}=n{\delta_{\alpha}}^{\beta}\bar{R}_i^{(n-1)}, &&
 \{\bar{Q}^{\alpha},\bar{L}_i^{(n)\beta}\}=i\epsilon^{\alpha\beta}C_i^{(n)}+n\epsilon^{\alpha\gamma}P_{i\gamma}^{(n-1)\beta},
\nonumber\\[2pt]
&
\{\bar{S}^{\alpha},\bar{L}_i^{(n)\beta}\}=-i\epsilon^{\alpha\beta}C_i^{(n+1)}-(n-2l+1)\epsilon^{\alpha\gamma}P_{i\gamma}^{(n)\beta}, && [I_{+},\bar{L}_i^{(n)\alpha}]=-\epsilon^{\alpha\beta}L_{i\beta}^{(n)},
\nonumber\\[2pt]
&
\{S_{\alpha},\bar{L}_i^{(n)\beta}\}=-{\delta_{\alpha}}^{\beta}(n-2l+1)\bar{R}_i^{(n)}, && [I_3,\bar{L}^{(n)\alpha}_i]=\frac{i}{2}\bar{L}_i^{(n)\alpha},
\nonumber\\[2pt]
&
[H,P_{i\alpha}^{(n)\beta}]=nP_{i\alpha}^{(n-1)\beta}, && [D,P_{i\alpha}^{(n)\beta}]=(n-l+1)P_{i\alpha}^{(n)\beta},
\nonumber\\[2pt]
&
[K,P_{i\alpha}^{(n)\beta}]=(n-2l+2)P_{i\alpha}^{(n+1)\beta}, && [Q_{\alpha},P_{i\beta}^{(n)\gamma}]=i\epsilon_{\alpha\beta}\bar{L}_i^{(n)\gamma}-n{\delta_{\alpha}}^{\gamma}Z_{i\beta}^{(n-1)},
\nonumber
\end{align}
\begin{align}
&
[\mathcal{J}_a,P_{i\alpha}^{(n)\beta}]=\frac{i}{2}{(\sigma_a)_\alpha}^{\gamma}P_{i\gamma}^{(n)\beta}-\frac{i}{2}P_{i\alpha}^{(n)\gamma}{(\sigma_a)_{\gamma}}^{\beta}, &&
[\bar{Q}^{\alpha},P_{i\beta}^{(n)\gamma}]=-i\epsilon^{\alpha\gamma}L_{i\beta}^{(n)}+n{\delta_{\beta}}^{\alpha}\bar{Z}_i^{(n-1)\gamma},
\nonumber\\[2pt]
&
[S_{\alpha},P_{i\beta}^{(n)\gamma}]=-i\epsilon_{\alpha\beta}\bar{L}_i^{(n+1)\gamma}+(n-2l+2){\delta_{\alpha}}^{\gamma}Z_{i\beta}^{(n)}, && [I_{-},P_{i\alpha}^{(n)\beta}]={\delta_{\alpha}}^{\beta}\bar{R}_i^{(n)},
\nonumber\\[2pt]
&
[\bar{S}^{\alpha},P_{i\beta}^{(n)\gamma}]=i\epsilon^{\alpha\gamma}L_{i\beta}^{(n+1)}-(n-2l+2) {\delta_{\beta}}^{\alpha} \bar{Z}_i^{(n)\gamma}, && [I_{+},P_{i\alpha}^{(n)\beta}]={\delta_{\alpha}}^{\beta}R_i^{(n)},
\nonumber\\[2pt]
&
[H,R_i^{(n)}]=nR_i^{(n-1)}, && [D,R_i^{(n)}]=(n-l+1)R_i^{(n)},
\nonumber\\[2pt]
&
[K,R_i^{(n)}]=(n-2l+2)R_i^{(n+1)}, && [Q_{\alpha},R_i^{(n)}]=-i L_{i\alpha}^{(n)}+n\epsilon_{\alpha\beta}\bar{Z}_i^{(n-1)\beta},
\nonumber\\[2pt]
&
[S_{\alpha},R_i^{(n)}]=iL_{i\alpha}^{(n+1)}-(n-2l+2)\epsilon_{\alpha\beta} \bar{Z}_i^{(n)\beta}, && [I_{-},R_i^{(n)}]=-P_{i\alpha}^{(n)\alpha},
\nonumber\\[2pt]
&
[I_3,R_i^{(n)}]=-iR_i^{(n)}, && [H,\bar{R}_i^{(n)}]=n\bar{R}_i^{(n-1)},
\nonumber\\[2pt]
&
[D,\bar{R}_i^{(n)}]=(n-l+1)\bar{R}_i^{(n)}, && [K,\bar{R}_i^{(n)}]=(n-2l+2)\bar{R}_i^{(n+1)},
\nonumber\\[2pt]
&
[\bar{Q}^{\alpha},\bar{R}_i^{(n)}]=-i\bar{L}_i^{(n)\alpha}+n\epsilon^{\alpha\beta}Z_{i\beta}^{(n-1)}, && [I_{+},\bar{R}_i^{(n)}]=-P_{i\alpha}^{(n)\alpha},
\nonumber\\[2pt]
&
[\bar{S}^{\alpha},\bar{R}_i^{(n)}]=i\bar{L}_i^{(n+1)\alpha}-(n-2l+2)\epsilon^{\alpha\beta}Z_{i\beta}^{(n)}, && [I_3,\bar{R}_i^{(n)}]=i\bar{R}_i^{(n)},
\nonumber\\[2pt]
&
[H,Z^{(n)}_{i\alpha}]=nZ_{i\alpha}^{(n-1)}, && [D,Z_{i\alpha}^{(n)}]=\left(n-l+\frac 32\right)Z_{i\alpha}^{(n)},
\nonumber\\[2pt]
&
[K,Z_{i\alpha}^{(n)}]=(n-2l+3)Z_{i\alpha}^{(n+1)}, && [\mathcal{J}_a,Z_{i\alpha}^{(n)}]=\frac{i}{2}{(\sigma_a)_{\alpha}}^{\beta}Z_{i\beta}^{(n)},
\nonumber\\[2pt]
&
\{Q_{\alpha},Z_{i\beta}^{(n)}\}=i\epsilon_{\alpha\beta}\bar{R}_i^{(n)}, && \{\bar{Q}^{\alpha},Z_{i\beta}^{(n)}\}=iP_{i\beta}^{(n)\alpha}+n{\delta_{\beta}}^{\alpha}W_i^{(n-1)},
\nonumber\\[2pt]
&
\{S_{\alpha},Z^{(n)}_{i\beta}\}=-i\epsilon_{\alpha\beta}\bar{R}_i^{(n+1)}, && [I_{+},Z_{i\alpha}^{(n)}]=-\epsilon_{\alpha\beta}\bar{Z}_i^{(n)\beta},
\nonumber\\[2pt]
&
\{\bar{S}^{\alpha},Z_{i\beta}^{(n)}\}=-iP_{i\beta}^{(n+1)\alpha}-(n-2l+3){\delta_{\beta}}^{\alpha}W_i^{(n)}, && [I_3,Z_{i\alpha}^{(n)}]=\frac{i}{2}Z_{i\alpha}^{(n)},
\nonumber\\[2pt]
&
[H,\bar{Z}^{(n)\alpha}_{i}]=n\bar{Z}_i^{(n-1)\alpha}, && [D,\bar{Z}_i^{(n)\alpha}]=\left(n-l+\frac 32\right)\bar{Z}_i^{(n)\alpha},
\nonumber\\[2pt]
&
[K,\bar{Z}_i^{(n)\alpha}]=(n-2l+3)\bar{Z}_i^{(n+1)\alpha}, && [\mathcal{J}_a,\bar{Z}_i^{(n)\alpha}]=-\frac{i}{2}\bar{Z}_i^{(n)\beta}{(\sigma_a)_{\beta}}^{\alpha},
\nonumber\\[2pt]
&
\{Q_{\alpha},\bar{Z}_i^{(n)\beta}\}=-i P_{i\alpha}^{(n)\beta}+n{\delta_{\alpha}}^{\beta}W_i^{(n-1)}, &&  \{\bar{Q}^{\alpha},\bar{Z}_i^{(n)\beta}\}=i\epsilon^{\alpha\beta}R_i^{(n)},
\nonumber\\[2pt]
&
\{S_{\alpha},\bar{Z}_i^{(n)\beta}\}=iP_{i\alpha}^{(n+1)\beta}-(n-2l+3){\delta_{\alpha}}^{\beta}W_i^{(n)}, && \{\bar{S}^{\alpha},\bar{Z}_i^{(n)\beta}\}=-i\epsilon^{\alpha\beta}R_i^{(n+1)},
\nonumber\\[2pt]
&
[I_{-},\bar{Z}_i^{(n)\alpha}]=\epsilon^{\alpha\beta}Z_{i\beta}^{(n)}, && [I_3,\bar{Z}_i^{(n)\alpha}]=-\frac{i}{2}\bar{Z}_i^{(n)\alpha},
\nonumber\\[2pt]
&
[H,W_i^{(n)}]=nW_i^{(n-1)}, && [D,W_i^{(n)}]=(n-l+2)W_i^{(n)},
\nonumber\\[2pt]
&
[K,W_i^{(n)}]=(n-2l+4)W_i^{(n+1)}, &&  [Q_{\alpha},W_i^{(n)}]=-iZ_{i\alpha}^{(n)},
\nonumber\\[2pt]
&
[\bar{Q}^{\alpha},W_i^{(n)}]=-i\bar{Z}_i^{(n)\alpha}, &&  [S_{\alpha},W_i^{(n)}]=iZ_{i\alpha}^{(n+1)},
\nonumber\\[2pt]
&
[\bar{S}^{\alpha},W_i^{(n)}]=i\bar{Z}_i^{(n+1)\alpha}. &&
\end{align}
In obtaining the structure relations (\ref{algebra}), the spinor algebra and properties of the Pauli matrices exposed in Appendix A have been used extensively. Above we omitted the standard $so(d)$ subalgebra generated by $M_{ij}$ and a chain of relations indicating that both $C_k^{(n)}$ and the generators in (\ref{extra1}) belong to the vector representation of $so(d)$: $[M_{ij},C_k^{(n)}]=-\delta_{ik} C_j^{(n)}+\delta_{jk} C_i^{(n)}$ etc.
Note that the range of values of the upper index $n$ which appears in braces in Eqs. (\ref{extra}) and  (\ref{extra1}) is unambiguously fixed by the structure of (anti)commutators involving the vector generators and $K$, $S_{\alpha}$, $\bar{S}^{\alpha}$ and the requirement that the superalgebra is finite--dimensional.

\vspace{0.5cm}

\noindent
{\bf 5. Conclusion}\\

To summarize, in this work we established the structure relations an $N=4$ supersymmetric extension of the $l$--conformal Galilei algebra based upon $osp(4|2)$ and provided its realization in terms of differential operators in superspace. This was achieved by combining the generators of spatial symmetries from the $l$--conformal Galilei algebra and those underlying the most general superconformal group in one dimension $D(2,1;\alpha)$. The value of the parameter $\alpha$ was fixed from the requirement that the resulting superalgebra was finite--dimensional. The analysis revealed $\alpha=-\frac 12$ thus reducing $D(2,1;\alpha)$ to $OSp(4|2)$.
In order to close the full superalgebra, a chain of extra generators carrying vector indices was introduced. These were interpreted as the fermionic partners of the acceleration generators $C_i^{(n)}$.

This work can be extended in several directions. Having fixed the structure relations of the superalgebra, one can proceed to the construction of dynamical systems enjoying the symmetry. Both second order and higher derivative models can be constructed by applying methods previously developed for the $l$--conformal Galilei group. Regarding models in nonrelativistic spacetime with cosmological constant, it would be interesting to construct the Newton--Hooke counterpart of the superalgebra in this work. The structure of admissible central extensions is worth studying as well. An intriguing open problem is to systematically develop a superfield description of models with $N=2$ and $N=4$ $l$--conformal Galilei supersymmetry.
A more speculative issue is a possible link between many-body models with $N=4$ $l$--conformal Galilei supersymmetry and a microscopic description of the near horizon extreme Myers--Perry black hole.

\vspace{0.5cm}

\noindent{\bf Acknowledgements}\\

\noindent
The work of I.M. was supported by the RF Presidential grant MK-2101.2017.2.

\noindent

\vspace{0.5cm}

\noindent
{\bf Appendix A}

\vspace{0.5cm}
\noindent
Throughout the text $SU(2)$--spinor
indices are raised and lowered with the use of the invariant
antisymmetric matrices
\be
\theta^\a=\e^{\a\b}\theta_\b, \quad {\bar\theta}_\a=\e_{\a\b} {\bar\theta}^\b,
\nonumber
\ee
where $\e_{12}=1$, $\e^{12}=-1$. Introducing the notation for the spinor bilinears
\be
\theta^2=\theta^\a \theta_\a, \qquad
\bar\theta^2=\bar\theta_\a \bar\theta^\a, \qquad \bar\theta \theta=\bar\theta^\a \theta_\a,
\nonumber
\ee
one gets
\be
\theta_\a \theta_\b=\frac 12 \e_{\a\b} \theta^2, \qquad \theta_\a \bar\chi_\b-\theta_\b \bar\chi_\a=\e_{\a\b} (\bar\chi \theta),
\nonumber
\ee
\be
\bar\theta^\a \bar\theta^\b=\frac 12 \e^{\a\b} \bar\theta^2,  \qquad \theta^\alpha {\bar\chi}^\beta-\theta^\beta {\bar\chi}^\alpha=-\epsilon^{\alpha \beta} (\bar\chi\theta).
\nonumber
\ee
The Pauli matrices ${{(\s_a)}_\a}^\b$
are chosen in the form
\be
\s_1=\begin{pmatrix}0 & 1\\
1 & 0
\end{pmatrix}\ , \qquad \s_2=\begin{pmatrix}0 & -i\\
i & 0
\end{pmatrix}\ ,\qquad
\s_3=\begin{pmatrix}1 & 0\\
0 & -1
\end{pmatrix}\ ,
\nonumber
\ee
which obey
\bea
&&
{{(\s_a \s_b)}_\a}^\b +{{(\s_b \s_a)}_\a}^\b=2 \d_{ab} {\d_\a}^\b \ , \quad
{{(\s_a \s_b)}_\a}^\b -{{(\s_b \s_a)}_\a}^\b=2i \e_{abc} {{(\s_c)}_\a}^\b \ ,
\nonumber\\[2pt]
&&
{{(\s_a \s_b)}_\a}^\b=\d_{ab} {\d_\a}^\b +i \e_{abc} {{(\s_c)}_\a}^\b \ , \quad
{{(\s_a)}_\a}^\b {{(\s_a)}_\g}^\r=2 {\d_\a}^\r {\d_\g}^\b-{\d_\a}^\b {\d_\g}^\r\ ,
\nonumber\\[2pt]
&&
{{(\s_a)}_\a}^\b \e_{\b\g} ={{(\s_a)}_\g}^\b \e_{\b\a}\ , \quad \e^{\a\b} {{(\s_a)}_\b}^\g=\e^{\g\b} {{(\s_a)}_\b}^\a \ ,
\nonumber
\eea
where $\e_{abc}$ is the totally antisymmetric tensor, $\e_{123}=1$. Our conventions for complex conjugation read
\bea
&&
{(\theta_\alpha)}^{*}=\bar\theta^\alpha, \qquad {\left(\bar\theta_\alpha\right)}^{*}=-\theta^\alpha, \qquad
{\left(\theta^2\right)}^{*}=\bar\theta^2, \qquad {\left(\bar\theta \sigma_a \chi\right)}^{*}=\bar\chi \sigma_a \theta.
\nonumber
\eea

\noindent

\vspace{0.5cm}

\noindent
{\bf Appendix B}

\vspace{0.5cm}
\noindent
The structure relations of the Lie superalgebra corresponding to $D(2,1;\alpha)$--supergroup read
\begin{align}\label{algebra}
&
[H,D]=H, && [H,K]=2D,
\nonumber\\[2pt]
&
[D,K]=K, && [\mathcal{J}_a,\mathcal{J}_b]=\epsilon_{abc} \mathcal{J}_c,
\nonumber\\[2pt]
&
\{ Q_\a, \bar Q^\b \}=-2 i H {\d_\a}^\b, &&
\{ Q_\a, \bar S^\b \}=-2\alpha {{(\s_a)}_\a}^\b \mathcal{J}_a+2iD {\d_\a}^\b+2(1+\alpha)I_3 {\d_\a}^\b,
\nonumber\\[2pt]
&
\{ S_\a, \bar S^\b \}=-2i K {\d_\a}^\b, &&
\{ \bar Q^\a, S_\b \}=2\alpha{{(\s_a)}_\b}^\a \mathcal{J}_a+2iD {\d_\b}^\a-2(1+\alpha)I_3 {\d_\b}^\a,
\nonumber\\[2pt]
&
\{ Q_\a, S_\b \}=2i (1+\alpha) \epsilon_{\alpha \beta} I_{-}, &&
\{ {\bar Q}^\a, {\bar S}^\b \}=-2i (1+\alpha) \epsilon^{\alpha \beta} I_{+},
\nonumber\\[2pt]
& [D,Q_\a] = -\frac{1}{2} Q_\a, && [D,S_\a] =\frac{1}{2} S_\a,
\nonumber\\[2pt]
&
[K,Q_\a] =S_\a, && [H,S_\a]=-Q_\a,
\nonumber\\[2pt]
&
[\mathcal{J}_a,Q_\a] =\frac{i}{2} {{(\s_a)}_\a}^\b Q_\b, && [\mathcal{J}_a,S_\a] =\frac{i}{2} {{(\s_a)}_\a}^\b S_\b,
\nonumber\\[2pt]
& [D,\bar Q^\a] =-\frac{1}{2} \bar Q^\a, && [D,\bar S^\a]=\frac{1}{2} \bar S^\a,
\nonumber\\[2pt]
& [K,\bar Q^\a] =\bar S^\a, && [H,\bar S^\a]=-\bar Q^\a,
\nonumber
\end{align}
\begin{align}
&
[\mathcal{J}_a,\bar Q^\a] =-\frac{i}{2} \bar Q^\b {{(\s_a)}_\b}^\a, && [\mathcal{J}_a,\bar S^\a]=-\frac{i}{2}
\bar S^\b {{(\s_a)}_\b}^\a,
\nonumber\\[2pt]
&
[I_{-},\bar Q^\a ]=\epsilon^{\alpha \beta} Q_\beta, && [I_{-},\bar S^\a ]=\epsilon^{\alpha \beta} S_\beta,
\nonumber\\[2pt]
&
[I_{+},Q_\a] =-\epsilon_{\alpha\beta} \bar Q^\b, && [I_{+},S_\a] =-\epsilon_{\alpha\beta} \bar S^\b,
\nonumber\\[2pt]
&
[I_3,Q_\a ]=\frac{i}{2} Q_\a, &&  [I_3,S_\a]=\frac{i}{2} S_\a,
\nonumber\\[2pt]
&
[I_3,\bar Q^\a]=-\frac{i}{2} \bar Q^\a, &&  [I_3,\bar S^\a] =-\frac{i}{2} \bar S^\a,
\nonumber\\[2pt]
&
[I_{-},I_3]=-i I_{-}, &&  [I_{+},I_3] =i I_{+},
\nonumber\\[2pt]
&
[I_{-},I_{+}]=2 i I_{3}. &&
\nonumber
\end{align}


\begin{thebibliography}{nn}
\bibitem{NOR}
J. Negro, M.A. del Olmo, A. Rodriguez-Marco, {\it Nonrelativistic conformal groups}, J. Math. Phys. {\bf 38} (1997) 3786.
\bibitem{H}
M. Henkel, {\it Local scale invariance and strongly anisotropic equilibrium critical systems}, Phys. Rev. Lett. {\bf 78} (1997) 1940, cond-mat/9610174.
\bibitem{FIL}	
S. Fedoruk, E. Ivanov, O. Lechtenfeld, {\it Superconformal mechanics}, J. Phys. A {\bf 45} (2012) 173001, arXiv:1112.1947.
\bibitem{G2}
A. Galajinsky, {\it Conformal mechanics in Newton--Hooke spacetime}, Nucl. Phys. B {\bf 832} (2010) 586, arXiv:1002.2290.
\bibitem{FL}
S. Fedoruk, J. Lukierski, {\it The algebraic structure of Galilean superconformal symmetries}, Phys. Rev. D
{\bf 84} (2011) 065002, arXiv:1105.3444.
\bibitem{PH}
P.A. Horv\'athy, {\it Non--relativistic conformal and supersymmetries}, Int. J. Mod. Phys. A {\bf 3} (1993) 339, arXiv:0807.0513.
\bibitem{DH}
C. Duval, P.A. Horv\'athy, {\it On Schrodinger superalgebras}, J. Math. Phys. {\bf 35} (1994) 2516, hep-th/0508079.
\bibitem{HU}
M. Henkel, J. Unterberger, {\it Supersymmetric extensions of Schrodinger--invariance}, Nucl. Phys. B {\bf 746} (2006) 155, math-ph/0512024.
\bibitem{AL} J.A. de Azcarraga, J. Lukierski, {\it Galilean superconformal symmetries}, Phys. Lett. B {\bf 678} (2009) 411, arXiv:0905.0141.
\bibitem{MS}
M. Sakaguchi, {\it Super  Galilean  conformal  algebra  in  AdS/CFT}, J. Math. Phys. {\bf 51} (2010) 042301, arXiv:0905.0188.
\bibitem{GLP}
A. Galajinsky, O. Lechtenfeld, K. Polovnikov, {\it N=4 superconformal Calogero models}, JHEP {\bf 0711} (2007) 008, arXiv:0708.1075.
\bibitem{FSS}
L. Frappat, P. Sorba, A. Sciarrino, {\it Dictionary on Lie superalgebras}, hep-th/9607161.
\bibitem{KL}
S. Krivonos, O. Lechtenfeld, {\it Many--particle mechanics with $D(2,1;\alpha)$ superconformal symmetry}, JHEP {\bf 1102} (2011) 042
arXiv:1012.4639.
\end{thebibliography}
\end{document}